\documentclass[11pt]{article}

\usepackage{amsmath}
\usepackage{amssymb}
\usepackage{amsmath}
\usepackage{amssymb}
\usepackage{amscd}
\newtheorem{Definition}{Definition}[section]
\newtheorem{Lemma}[Definition]{Lemma}
\newtheorem{Theorem}[Definition]{Theorem}

 \newcommand{\beq}{\begin{equation}}
\newcommand{\eeq}{\end{equation}} 
\newcommand{\bea}{\begin{eqnarray}}
\newcommand{\eea}{\end{eqnarray}} \newcommand{\nn}{\nonumber}

 \newcommand{\qm}{quantum
mechanics}

\newcommand{\Hs}{Hilbert space}

\newcommand{\raw}{\rightarrow}

\newcommand{\LRaw}{\Leftrightarrow}
 
\newcommand{\ot}{\otimes} 
\newcommand{\la}{\langle} \newcommand{\ra}{\rangle}

\newcommand{\half}{\mbox{\footnotesize $\frac{1}{2}$}}

\newcommand{\inv}{^{-1}}

\newcommand{\er}{\eqref}
\newcommand{\al}{\alpha}

 \newcommand{\varep}{\varepsilon}

 \newcommand{\kp}{\kappa}
\newcommand{\lm}{\lambda} \newcommand{\Lm}{\Lambda}
 \newcommand{\sg}{\sigma}
  
 \newcommand{\phv}{\varphi}
 \newcommand{\ps}{\psi} 
 
\newcommand{\CT}{{\mathcal T}}

\newcommand{\C}{{\mathbb C}} 
 
\newcommand{\N}{{\mathbb N}} \newcommand{\R}{{\mathbb R}}
 
  \makeatletter
\newskip\tempskip \def\endproof{{\parfillskip24\p@ plus\@ne
fil\@@par}\tempskip\prevdepth
\ifdim\lastskip=\z@\tempskip\z@\else\vskip-\lastskip
\ifdim\tempskip>4\p@ \tempskip.5\tempskip \else \tempskip\z@\fi\fi
\nobreak\vskip-\baselineskip\vskip-\tempskip\noindent\hbox
to\hsize{\hfill
$\blacksquare$}\par\vskip\tempskip\vskip\abovedisplayskip\@doendpe}
\makeatother \makeatletter
\newskip\tempskip \def\endiproof{{\parfillskip24\p@ plus\@ne
fil\@@par}\tempskip\prevdepth
\ifdim\lastskip=\z@\tempskip\z@\else\vskip-\lastskip
\ifdim\tempskip>4\p@ \tempskip.5\tempskip \else \tempskip\z@\fi\fi
\nobreak\vskip-\baselineskip\vskip-\tempskip\noindent\hbox
to\hsize{\hfill
$\Box$}\par\vskip\tempskip\vskip\abovedisplayskip\@doendpe}
\makeatother 

\newcommand{\hv}{hidden variable}

\topmargin = - 0.5 cm \textheight = 23 cm \textwidth = 15 cm
\begin{document} 
\pagenumbering{arabic} \setlength{\unitlength}{1cm}\cleardoublepage
\thispagestyle{empty}
\title{On the Colbeck--Renner Theorem}
\author{Klaas Landsman\\ \mbox{} \hfill \\
Institute for Mathematics, Astrophysics, and Particle Physics\\ Faculty of Science, Radboud University Nijmegen\\
\texttt{landsman@math.ru.nl}}
\date{\today}
\maketitle
 \begin{abstract} 
\noindent 
In three papers Colbeck and Renner (\emph{Nature Communications} 2:411, (2011); \emph{Phys.\ Rev.\ Lett.}  108, 150402 (2012);  \texttt{arXiv:1208.4123}) argued that ``no alternative theory compatible with quantum theory and satisfying the freedom of choice assumption can give improved predictions.'' We give a more precise version of the formulation and proof of this remarkable claim. Our proof broadly follows theirs, which relies on physically well motivated axioms,  but to fill in some crucial details  certain technical assumptions have had to be added,
whose  physical status seems somewhat obscure.
\end{abstract}
\maketitle
\section{Introduction}
The  claim by Colbeck and Renner  that  ``no alternative theory compatible with quantum theory and satisfying the freedom of choice assumption can give improved predictions'' \cite{CR1,CR2,CR3} has attracted considerable attention (see e.g. the review \cite{Leifer}),  some of which has been rather  critical \cite{GR,Laudisa,Leegwater}. 
The aim of this paper is to give a watertight proof of their theorem, including a statement of precise, mathematically formulated assumptions. 

Our proof broadly follows the dazzling reasoning of Colbeck and Renner, except that some of their theoretical physics style heuristic arguments have been replaced by rigorous mathematics. However, if this had been a routine exercise in mathematical physics we would not have taken the effort. The point of our analysis is to show that additional assumptions are necessary to make the proof work, so that the theorem is weaker than it may appear to be at first sight: it does not show that \qm\ is complete, but that (informative) extensions are subject to (possibly undesirable) constraints. 

Indeed, apart from three
physically natural (and unavoidable) assumptions, namely  \emph{Compatibility with Quantum Mechanics},
 \emph{Parameter Independence} (the latter being a well-known \hv\ version of the no-signaling axiom), and what we call
  \emph{Product Extension}, 
  we also need three
assumptions that are satisfied by \qm\ itself but might seem somewhat unnatural if imposed on a \hv\ theory, viz.\ \emph{Continuity of Probabilities},
 \emph{Unitary Invariance}, and what we call \emph{Schmidt Extension}.
 We also replaced the original probabilistic setting, in which almost everything (including even the quantum state) was treated as a random variable, by a more conventional \hv\ theory perspective (which  circumvents some unnecessary controversies \cite{GR,SU}). Our approach differs significantly from  interesting recent work of Leegwater \cite{Leegwater}, which has a similar goal. 
  \section{Notation}\label{Not}
A \emph{\hv\ theory}  $\mathcal{T}$ underlying \qm\  yields probabilities   $$P(Z_1=z_1,\ldots, Z_n=z_n|\lm)\equiv P(\vec{Z}=\vec{z}|\lm)$$ for the possible outcomes $\vec{z}=(z_1,\ldots, z_n)$ of a measurement of any family $\vec{Z}=(Z_1,\ldots, Z_n)$ of
 commuting hermitian operators  on any \Hs\ $H$ (here assumed to be finite dimensional for simplicity), given an arbitrary parameter $\lm\in\Lm$ (i.e., the `\hv'), where $\Lm$ is some Borel space.\footnote{This generality, which is not a common feature of hidden variable theories (and as such is already a significant assumption), is necessary for the Colbeck--Renner argument to work.}
 Being `classical' probabilities, these numbers are \emph{a priori} only supposed to satisfy
$0\leq P(\vec{Z}=\vec{z}|\lm) \leq 1$ and
$\sum_{\vec{z}}  P(\vec{Z}=\vec{z}|\lm)=1$,
where the sum is over all possible outcomes. It will follow from the assumptions below that necessarily $z_i\in \sg(Z_i)$ (i.e., the spectrum of $Z_i$) for each $i=1,\ldots, n$, in the sense that $P(\vec{Z}=\vec{z}|\lm)=0$ if this is not the case. 
Families of operators $\vec{Z}_c$ (all defined on the same $H$) are indexed by some parameter $c\in C$, called the ``setting'' of the experiment.\footnote{Colbeck and Renner look at the setting $c$ as the value of some random variable $C$, but this is controversial \cite{SU}; for us, $C$ is simply the set in which $c$ takes values.} 

An important special case will be the bipartite setting  $H=H_1\ot H_2$, where Alice and Bob measure  hermitian operators $X$ and $Y$ on $H_1$ and $H_2$, repectively, so that $n=2$, $Z_1=X\ot 1_{H_2}$ and $Z_2=1_{H_1}\ot Y$. We then write $z_1=x$, $z_2=y$, and $c=(a,b)$, so that we typically look at expressions like
 $P(X_a=x,Y_b=y|\lm)$.
The other case of interest will simply be $n=1$ with $Z_1\equiv Z$, $z_1\equiv z$; indeed, this will be the case in the statement of the theorem (the bipartite case playing a role only in the proof, though a crucial one!).

In this paper,  quantum-mechanical states will just be unit vectors $\psi\in H$. The corresponding prediction for the above probabilities, i.e., the `Born rule', is given by \cite{vN32}
\begin{equation}
P_{\psi}(\vec{Z}=\vec{z})= \la\ps, E_{\vec{Z}}(\vec{z})\ps\ra, \label{BR}
\end{equation}
where $ E_{\vec{Z}}(\vec{z})=\prod_{i=1}^n E_{Z_i}(z_i)$, in which $E_{Z_i}(z_i)$ is the spectral projection on the eigenspace $H_{z_i}\subset H$ of $Z_i$
(i.e., $Z_i\psi=z_i\psi$ iff $\psi\in H_{z_i}$). 
As detailed in \S\ref{assec}, $\CT$ assigns a probability measure $\mu_{\psi}$ on $\Lm$ to each state $\ps$. The following notation occurs throughout the paper:
\begin{equation}
P_{\psi}(\vec{Z}=\vec{z}|\lm)=\al(\lm), \label{al1}
\end{equation}
with $\al:\Lm\raw[0,1]$ an explicitly given measurable function (often constant). This 
means:\footnote{Colbeck and Renner treat  $\psi$ as a random variable and hence interpret 
$P_{\psi}(\vec{Z}=\vec{z}|\lm)$ as a  probability conditioned on knowing (that) $\psi$. We do not do so, yet our mathematical unfolding of \er{al1} is similar.}
\begin{quote}
$P(\vec{Z}=\vec{z}|\lm)=\al(\lm)$  for almost every  $\lm$ with respect to the measure 
$\mu_{\ps}$.\footnote{In other words, there is a subset $\Lm'\subset\Lm$ such that $\mu_{\ps}(\Lm')=0$ and
$P_{\psi}(\vec{Z}=\vec{z}|\lm)=\al(\lm)$
holds for any $\lm\in\Lm\backslash\Lm'$. If $\Lm$ is finite, this simply means that  the equality holds for any $\lm$ for which $\mu_{\ps}(\{\lm\})>0$.}
\end{quote} 
 Since this notation renders equalities like
 \begin{equation}
P_{\psi}(\vec{Z}=\vec{z}|\lm)=P_{\phv}(\vec{Z}' =\vec{z}'|\lm), \label{al2}
\end{equation}
ambiguous (where $\ps,\phv$ are states in $H$), we explicitly define \er{al2} as the
 double implication $$P_{\psi}(\vec{Z}=\vec{z}|\lm)=\al(\lm\:)\Leftrightarrow\: P_{\phv}(\vec{Z}'=\vec{z}'|\lm)=\al(\lm).$$
 
This notation also appears in  our final pair of conventions: for $\varep\raw 0$ we write
\begin{eqnarray}
\psi \stackrel{\varep}{\approx}\phv &\LRaw &(1-\varep)\leq |\la\ps,\phv\ra|\leq 1; \label{epnot}\\
P_{\ps}(\vec{Z}=\vec{z}|\lm) \stackrel{\varep}{\approx}P_{\phv}(\vec{Z}'=\vec{z}'|\lm) &\LRaw &
P_{\ps}(\vec{Z}=\vec{z}|\lm)=P_{\phv}(\vec{Z}'=\vec{z}'|\lm) + O(\sqrt{\varep}).
\end{eqnarray}
\section{Assumptions}\label{assec}
The assumptions in our reformulation of the Colbeck--Renner Theorem are as follows. 
  \begin{description}  
\item[CQ] \emph{Compatibility with Quantum Mechanics:}  for any unit vector $\ps\in H$, the theory
 $\mathcal{T}$ yields a  state $\mu_{\psi}$  (i.e., a probability measure on $\Lm$),\footnote{As the notation indicates, $\mu_{\psi}$ depends on $\psi$ only and hence is independent of $Z$ and $z$. From the point of view of  $\mathcal{T}$, a quantum state \emph{is} a probability measure on $\Lm$, so one might even write $\psi$ for $\mu_{\psi}$.}  such that (cf.\ \er{BR})
 \begin{equation}
\int_{\Lm}d\mu_{\psi}(\lm)\, P(\vec{Z}=\vec{z}|\lm)=P_{\psi}(\vec{Z}=\vec{z}). \label{cqm}
\end{equation}
\item[UI] \emph{Unitary Invariance:} for any unit vector $\ps\in H$ and unitary operator $U$ on $H$,\footnote{This assumption may be replaced by its main consequence, i.e., Lemma \ref{L1} below.}
\begin{equation}
P_{U\psi}(\vec{Z}=\vec{z}|\lm)=P_{\psi}(U\inv\vec{Z}U=\vec{z}|\lm).
\end{equation}
\item[CP] \emph{Continuity of Probabilities:} If $\psi \stackrel{\varep}{\approx}\phv$, then $P_{\ps}(\vec{Z}=\vec{z}|\lm) \stackrel{\varep}{\approx}P_{\phv}(\vec{Z}=\vec{z}|\lm)$.
 \end{description}
 In the remaining three axioms,  $H=H_1\ot H_2$, and  $X$ and $Y$ are hermitian operators on $H_1$ and $H_2$, respectively 
 (identified with operators $X\ot 1_{H_2}$ and $1_{H_1}\ot Y$ on $H$ as appropriate).
   \begin{description}  
\item[PI]  \emph{Parameter Independence:}\footnote{In words, this assumptions states that
 the probabilities for Alice's measurement outcomes, given $\lm$,  are not only independent of Bob's  choice of his observable $Y$, but are even independent of his existence altogether, as they are given by the expression that $\mathcal{T}$ yields for Alice's experiment alone (and likewise for Bob).
This slightly  generalizes the usual Parameter Independence in the context of Bell's Theorem  \cite{Bub}. Note that in our form \textbf{PI} only makes sense 
because  \er{BR} and \er{cqm} imply that  for $P_{\psi}(\vec{Z}=\vec{z}|\lm)$ to be nonzero (in the sense of \S\ref{Not}) we must have $z_i\in\sg(Z_i)$ for each $i$.}
\begin{eqnarray}
\sum_{y\in\sg(Y)}P(X=x,Y=y|\lm)&=&P(X=x|\lm);\\
\sum_{x\in\sg(X)}P(X=x,Y=y|\lm)&=&P(Y=y|\lm).
\end{eqnarray}
\item[PE] \emph{Product  Extension:}  for any pair of states $\psi_1\in H_1$,  $\psi_2\in H_2$,
\begin{equation}
P_{\psi_1}(X=x|\lm)=P_{\psi_1\ot\psi_2}(X=x|\lm). \label{L22}
\end{equation}
\item[SE] \emph{Schmidt  Extension:} if  $e_i\in H_1$ ($i=1,\ldots,\dim(H)$) are eigenstates  of $X$, then for arbitrary orthogonal states  $u_i\in H_2$ and arbitrary  coefficients $c_i>0$ with $\sum_i c^2_i=1$, 
\begin{equation}
P_{\sum_ic_i\cdot e_i}(X=x|\lm)=P_{\sum_i c_i\cdot  e_i\ot u_i}(X=x|\lm). \label{L21}
\end{equation}
 \end{description}
 \emph{Comments.} All assumptions are satisfied by \qm\ itself (seen as a `hidden' variable theory, with the state $\psi$ as the `hidden' variable $\lm$ \cite{BB}). 
 In the broader context of \hv\ theories,  \textbf{CQ} seems unavoidable in any such discussion, and also \textbf{PI} and
  \textbf{PE}   have convincing physical plausibility. Unfortunately,   the other assumptions are  purely technical and  have solely been invented to carry out certain  steps in the proof. 
  
 In particular, although  \textbf{UI},   \textbf{CP}, and \textbf{SE}  represent the  essence of \qm\ itself, these assumptions 
 are far from self-evident for a \hv\ theory. Moroever, the former two are quite unsatisfactory, in that they do not merely constrain the
probabilities  $P(\vec{Z}=\vec{z}|\lm)$ of $\CT$: they rather involve an interplay between these probabilities and the supports of the measures $\mu_{\ps}$ and $\mu_{U\ps}$. We challenge the reader to economize this!
\section{Theorem and proof}
 Our reformulation of the Colbeck--Renner Theorem, then, is as follows. 
\begin{Theorem}\label{CRT}
 If some \hv-theory $\mathcal{T}$ satisfies  \textbf{CQ},  \textbf{UI},  \textbf{CP}, \textbf{PI}, \textbf{PE}, and  \textbf{SE}, then 
 for any (finite-dimensional) \Hs\ $H$, state $\psi\in H$, and observable $Z$ on $H$,  
 \beq
 P_{\psi}(Z=z|\lm)=P_{\psi}(Z=z).\label{CR1}
 \eeq
\end{Theorem}
We first assume (without loss of generality) that $Z$ is nondegenerate as a hermitian matrix, in that it has  distinct eigenvalues $(z_1,\ldots, z_{\dim(H)})$. This assumption will be justified at the end of the proof.
The proof consists of three steps:
\begin{enumerate}
\item The theorem holds for $H=\C^2$ and any pair $(Z,\psi)$ for which 
\begin{equation}
P_{\psi}(Z=z_1)=P_{\psi}(Z=z_2)=1/2, \label{Step1eq}
\end{equation}
This only requires  assumptions \textbf{CQ},  \textbf{PI}, and  \textbf{SE}.
 \item The theorem holds for $H=\C^l$, $l<\infty$ arbitrary,  and any pair $(Z,\psi)$ for which 
\begin{equation}
P_{\psi}(Z=z_1)=\cdots =P_{\psi}(Z=z_l)=1/l. \label{Step2eq}
\end{equation}
This is just a slight extension of step 1 and uses the same three assumptions.
 \item The theorem holds in general. This requires  all assumptions (as well as step 2).
\end{enumerate}
The first step is mathematically straightforward but physically quite deep, depending on \emph{chained Bell inequalities} \cite{BC}, and is due to \cite{CR3} (we will give a slightly simplified proof below). 
The second step is easy. The third step, relying on the technique of \emph{embezzlement} \cite{vDH},  is highly nontrivial.
This is step that  our analysis mainly attempts to clarify. 
\subsection*{Step 1}
Let $H=\C^2$, with  basis $(e_1,e_2)$ of eigenvectors of $Z$, so that $\ps\in \C^2$ may be written as 
\beq\ps=(e_1+e_2)/\sqrt{2}.\eeq
Without loss of generality, we may assume that $z_1=1$ and $z_2=-1$. 
We now relabel $Z$ as $Z_0$ and extend it to a family of operators $(Z_k)_{k=0,1,\ldots, 2N-1}$ by fixing an integer $N>1$, putting  $\theta_k=k\pi/2N$, and defining
\begin{equation}
Z_k=[\theta_{k+\pi}]-[\theta_k] ,
\end{equation}
where, for any angle $\theta\in[0,2\pi]$, the operator 
$[\theta]=|\theta\rangle\langle\theta|$ is the orthogonal projection onto the subspace (ray) spanned by the unit vector
  \begin{equation}
|\theta\rangle=\sin(\theta/2)\cdot e_1+\cos(\theta/2)\cdot e_2.
\end{equation}
 In the corresponding bipartite setting, we have  observables $X_k\equiv Z_k\ot 1_2$ and $Y_k\equiv 1_2\ot Z_k$ on $\C^2\ot\C^2$, as well as a maximally correlated (Bell) state  $\psi_{AB}\in \C^2\ot\C^2$, given by
\begin{equation}
\psi_{AB}=\frac{1}{\sqrt{2}}(e_1\ot e_1+e_2\ot e_2). \label{psiepr2}
\end{equation}
Using  assumptions \textbf{PI} and \textbf{SE}, we then have, for $i=1,2$ $z_1=1$, and $z_2=-1$,
\begin{equation}
P_{\psi}(Z=z_i|\lm)=P_{\psi_{AB}}(X_0=z_i|\lm). \label{step11}
\end{equation}
The quantum-mechanical prediction is 
\beq
P_{\psi_{AB}}(X_0=1)=P_{\psi_{AB}}(X_0=-1)=\half.
\eeq
 As in \cite{CR3}, our goal is to show that also
\begin{equation}
P_{\psi_{AB}}(X_0=1|\lm)=P_{\psi_{AB}}(X_0=-1|\lm)=\half. \label{step12}
\end{equation}
To this effect we introduce the combination of probabilities
\begin{equation}
I^{(N)}(\lm)=P(X_0=Y_{2N-1}|\lm)+\sum_{a\in A_N,b\in B_N,|a-b|=1} P(X_a\neq Y_b|\lm),
\end{equation}
where $A_N= \{0,2,\ldots, 2N-2\}$ and $B_N= \{1,3,\ldots, 2N-1\}$.
The inequality  \cite{Leegwater}
\begin{eqnarray}
|P(X_a=x_i|{\lm})-P(Y_b=x_i|{\lm})| &=& |P(X_a=x_i, Y_b=x_i|{\lm})+P(X_a=x_i, Y_b\neq x_i|{\lm})\nn\\
&-& | P(X_a=x_i, Y_b=x_i|{\lm})-P(X_a\neq x_i, Y_b=x_i|{\lm})|\nn\\
&=& |P(X_a=x_i, Y_b\neq x_i|{\lm})-P(X_a\neq x_i, Y_b=x_i|{\lm})|\nn\\
&\leq & P(X_a=x_i, Y_b\neq x_i|{\lm})+P(X_a\neq x_i, Y_b=x_i|{\lm})\nn\\
&= & P(X_a\neq Y_b|{\lm}), 
\end{eqnarray}
where $i=1,2$, and we used \textbf{PI}, implies a further inequality: since $X_{2N}=-X_0$,
\begin{eqnarray*}
|P(X_0=1|{\lm})-P(X_0=-1|{\lm})|&=& |P(X_0=1|{\lm})-P(X_{2N}=1|{\lm})| \nn\\
&\leq &\sum_{a,b,|a-b|=1} |P(X_a=1|{\lm})-P(Y_b=1|{\lm})| \nn\\
&\leq &\sum_{a,b,|a-b|=1}  P(X_a\neq Y_b|{\lm})
\leq I^{(N)}({\lm}).
\end{eqnarray*}
Integrating this with respect to the measure $\mu_{\ps_{AB}}$ and using \textbf{CQ} gives
\begin{equation}
\int_{\Lm}d\mu_{\psi_{AB}}(\lm)\, |P(X_0=1|\lm)-P(X_0=-1|\lm)|\leq
\int_{\Lm}d\mu_{\psi_{AB}}(\lm)\,  I^{(N)}(\lm)=I^{(N)}_{\psi_{AB}}. \label{fineq}
\end{equation}
A routine calculation shows that 
 the  quantum-mechanical prediction $I^{(N)}_{\psi_{AB}}$ is given by 
\begin{equation}
I^{(N)}_{\psi_{AB}}=2N\sin^2(\pi/4N), 
\end{equation}
so that
\beq\lim_{N\raw\infty}I^{(N)}_{\psi_{AB}}=0.
\eeq
Letting $N\raw\infty$ in \er{fineq} therefore yields \er{step12}. From \er{step11} we then obtain  \er{Step1eq}.
 \subsection*{Step 2}
Let $H=\C^l$ and let $(e_i)_{i=1}^l$ be an orthonormal basis of eigenvectors of $Z$, with corresponding eigenvalues $z_i$, and phase factors for the eigenvectors $e_i$  such that $c_i>0$ in the expansion
\beq\psi=\sum_i c_ie_i.\label{psiei}
\eeq  Of course, $\sum_i c_i^2=1$. The case of interest will be $c_1=\cdots = c_l=1/l$, but first
we  merely assume that $c_1=c_2$ (the same reasoning applies to any other pair), with $z_1=1$ and $z_2=-1$ (which involves no loss of generality either and just simplifies the notation). The other coefficients $c_i$ ($i> 2$) may or may not be equal to $c_1$. 

Generalizing \er{step12}, we will show that
\begin{equation}
P_{\psi}(Z=1|\lm)=P_{\psi}(Z=-1|\lm). \label{step21}
\end{equation}
This shows that if two Born probabilities defined by some quantum state $\psi$ are equal, then the underlying \hv\ probabilities (conditioned on $\psi$) must be equal, too. 
Eq.\ \er{Step2eq} immediately follows from this result by taking all $c_i$  to be equal.

Given step 1, the derivation of \er{step21} is a piece of cake. We again pass to the bipartite setting, introducing  two copies $H_A=H_B=\C^l$ of $H$, and define the correlated state
 \begin{equation}
\psi_{AB}=\sum_i c_i \cdot e_i\ot e_i \label{psiepr3}
\end{equation}
in $H_A\ot H_B$. Eq.\ \er{step11} again follows from  assumptions \textbf{PI} and \textbf{SE}. Throughout the argument
 of step 1,  we now replace each probability $P(X_a=x,Y_b=y|\lm)$ by a corresponding probability $P^{(1)}(X_a=x,Y_b=y|\lm)$, defined as 
 the conditional probability 
\begin{eqnarray}
P^{(1)}(X_a=x,Y_b=y|\lm)&=&P(X_a=x,Y_b=y||x|=|y|=1,\lm)\nn \\
&=&\frac{P(X_a=x,Y_b=y,|x|=|y|=1|\lm)}{P(|x|=|y|=1|\lm)},
\end{eqnarray}
for all $\lm$ for which  $P(|x|=|y|=1|\lm)>0$, whereas 
\begin{equation}
P^{(1)}(X_a=x,Y_b=y|\lm)=0 
\end{equation}
whenever $P(|x|=|y|=1|\lm)=0$.
The same argument then yields \er{fineq}, with $P$ replaced by $P^{(1)}$ but with the same right-hand side;
see \cite[\S3.2]{Leegwater} for this calculation. As in step 1,
\beq
P^{(1)}_{\psi_{AB}}(X_0=1|\lm)=P^{(1)}_{\psi_{AB}}(X_0=-1|\lm), \label{cancel1}
\eeq
which implies that \beq
P_{\psi_{AB}}(X_0=1|\lm)=P_{\psi_{AB}}(X_0=-1|\lm), \label{step2f}
\eeq
either because both sides vanish (if $P(|x|=|y|=1|\lm)=0$), or because (in the opposite case) the denominator $P(|x|=|y|=1|\lm)$
cancels from both sides of \er{cancel1}. 

Combined with \er{step11}, eq.\  \er{step2f} proves \er{step21} and hence establishes step 2.
 \subsection*{Step 3}
We continue to use the notation established at the beginning of step 2, especially \er{psiei}. 
 As in step 1, we introduce two copies $H_A=H_B=\C^l$ of $H$, as well as two  states
\begin{eqnarray}
\ps_{AB}&=&\sum_i c_i \cdot e_i\ot e_i\in H_A\ot H_B;\\
\ps_{AB}'''&=& \kp_n\ot e_1'\ot e_1'\ot \ps_{AB}\in H_A'''\ot H_B''',
\end{eqnarray}
 where $\kp_n$ is given by \er{mun}, 
$H'''=H''\ot H'\ot H$, and we have notationally ignored the obvious permutations of factors in the tensor product.

For any $\varep>0$ and given coefficients $c_i$,  pick $c_i'$ in $\R^+$ such that $(c_i')^2\in\mathbb{Q}^+$ and 
\begin{equation}
|c_i'-c_i|<\varep/\dim(H),\label{old4.36}
\end{equation}
which implies that, in the sense of \er{epnot},
$\sum_i c_i' e_i \stackrel{\varep/2}{\approx}\sum_i c_i e_i$.
 Suppose $c_i'=\sqrt{p_i/q_i}$, with $p_i,q_i\in\N$ and $\mathrm{gcd}(p_i,q_i)=1$, and define 
\beq
m_i=p_i\prod_{i'\neq i}q_{i'}.\eeq
Consequently, writing $q=1/\sqrt{\sum_{i'}m_{i'}}$, the following quotient is independent of $i$:
\begin{equation}
\frac{c_i'}{\sqrt{m_i}}=q.\label{cmq}
\end{equation}
Given the integers $m_i$ thus obtained, we define a unitary operator $U:H'''\raw H'''$ by
\begin{equation}
U=\sum_{i=1}^l U^{(m_i)}\ot P_i,
\end{equation}
where $P_i:H\raw H$ projects onto $e_i$ (that is, $P_i=|e_i\ra\la e_i|$ in physics notation) and $U^{(m_i)}$ is defined in \er{defUi}. From this definition  (with additional labels to denote the copies $U_A:  H_A'''\raw H_A'''$ and $U_B:  H_B'''\raw H_B'''$)
and \er{fe} and \er{old4.36}, we then obtain the relations
\begin{eqnarray}
1_{H'''_A} \ot 1_{H'''_B}  (\ps_{AB}''')& =&\kp_n\ot  \sum_{i=1}^l c_i\cdot   \xi_{AA'}^{i1}\,\ot \xi_{BB'}^{i1};\label{key1}\\
U_A\ot 1_{H'''_B}  (\ps_{AB}''')& =& \frac{1}{\sqrt{C(n)}}\sum_{i=1}^l\sum_{k=1}^n\frac{c_i}{\sqrt{k}}\cdot e''_{s_k}\ot e''_k\, \ot  \xi_{AA'}^{ij^i_k}\,\ot \xi_{BB'}^{i1}
;\label{key2}\\
1_{H'''_A} \ot U_B (\ps_{AB}''')& =& \frac{1}{\sqrt{C(n)}}\sum_{i=1}^l\sum_{k=1}^n\frac{c_i}{\sqrt{k}}\cdot  e''_{k}\ot e''_{s_k}\, \ot  \xi_{AA'}^{i1}\,\ot \xi_{BB'}^{ij^i_k}
;\label{key3}\\
U_A\ot U_B (\ps_{AB}''')& \stackrel{\varep}{\approx}&q\cdot\kp_n\ot
\sum_{i=1}^l\sum_{j_{i}=1}^{m_i}\xi_{AA'}^{ij_i}\,\ot \xi_{BB'}^{ij_i}.
\label{key4}
\end{eqnarray}
Here
\begin{eqnarray}
\xi^{ij}&=&e_i\ot e_{j}'\in H\ot H',
\end{eqnarray}
with corresponding copies $\xi^{ij_i}_{AA'}\in H_A\ot H_A'$ and $\xi_{BB'}^{ij_i}\in H_B\ot H_B'$;  the right-hand sides of   \er{key1} - \er{key4}  have been arranged so as to obtain vectors in the six-fold tensor product $$H_A''\ot H_B''\ot H_A\ot H_A'\ot H_B\ot H_B'.$$ 
\noindent
 The following (sub)steps are meant to replace (or  justify) the core argument of \cite{CR3}. We repeatedly invoke the following lemma, whose proof just unfolds the notation (which incorporates the identification of $X$ with $X\ot 1_{H_2}$ and of $Y$ with $1_{H_1}\ot Y$  as appropriate).
\begin{Lemma}\label{L1}
Assume  \textbf{PI} and  \textbf{UI}. For any pair of unitary operators $U_1$ on $H_1$  and $U_2$ on $H_2$, and  any unit vector $\psi\in H_1\ot H_2$, one has
\begin{eqnarray}
P_{(U_1\ot 1_{H_2})\ps}(Y=y|\lm)&=&P_{\ps}(Y=y|\lm);\\
P_{(1_{H_1}\ot U_2)\ps}(X=x|\lm)&=& P_{\psi}(X=x|\lm).
\end{eqnarray}
\end{Lemma}

We now introduce some convenient notation. 
Since we assume that $Z$ is nondegenerate,  there is a bijective correspondence between its eigenvalues $Z=z_i$ and its eigenvectors $e_i$. Instead of $P(Z=z_i)$ dressed with whatever parameters $\psi$ or $\lm$, we may then write
$P(e_i)$, where $Z$ is understood, and analogously for the more complicated operators on tensor products of \Hs\ appearing  below. We are now in a position to go ahead:

\begin{itemize}
\item From Step 2, using the notation explained below \er{psiei}, 
\begin{equation}
P_{q\cdot \sum_{i=1}^l\sum_{j_{i}=1}^{m_i} \xi_{BB'}^{ij_i}}(\xi_{BB'}^{ij}|\lm)=q^2. \label{q1}
\end{equation}
\item From \er{L21} in \textbf{PE} and \er{q1},
\begin{equation}
P_{q\cdot \sum_{i,j_{i}} \xi_{AA'}^{ij_i}\,\ot\xi_{BB'}^{ij_i}}(\xi_{BB'}^{ij}|\lm)=q^2. \label{q2}
\end{equation}
\item From \er{L22} in \textbf{SE} and \er{q2},
\begin{equation}
P_{q\cdot\kp_n\ot
\sum_{i,j_{i}}\xi_{AA'}^{ij_i}\,\ot \xi_{BB'}^{ij_i}}(\xi_{BB'}^{ij}|\lm)=q^2. \label{q3}
\end{equation}
\item From \er{q3},  \textbf{CP} (whose notation we use), and \er{key4},
\begin{equation}
P_{U_A\ot U_B (\ps_{AB}''')}( \xi_{BB'}^{ij}|\lm)\stackrel{\varep}{\approx}q^2. \label{q4}
\end{equation}
\item From  \er{q4} and Lemma \ref{L1}, we have
\begin{equation}
P_{1_{H'''_A}\ot U_B (\ps_{AB}''')}( \xi_{BB'}^{ij_i}|\lm) \stackrel{\varep}{\approx}q^2\:\: ( j_i=1,\ldots, m_i), \label{q5a}
\end{equation}
whereas the definition of the indices in question gives
\begin{equation}
P_{1_{H'''_A}\ot U_B (\ps_{AB}''')}( \xi_{BB'}^{ij_i}|\lm) \stackrel{\varep}{\approx}0\:\: (j_i=m_i+1,\ldots, m); \label{q5b}
\end{equation}
here the number $m$ (satisfying $m\geq m_i$ for all $i$) is introduced in the Appendix.
\end{itemize}
We now start a different argument, to be combined with \er{q5a} - \er{q5b} in due course.
\begin{itemize}
\item From  \textbf{PE}, \textbf{SE}, and \er{psiei}, with $e_A^i\in H_A$ denoting $e_i\in H$, we have
\begin{equation}
P_{\psi}(Z=z_i|\lm)\equiv P_{\psi}(e_i|\lm)=P_{\kp_n\ot  \sum_i c_i\cdot   \xi_{AA'}^{i1}\,\ot \xi_{BB'}^{i1}}(e_A^i|\lm).\label{c1}
\end{equation}
\item Using Lemma \ref{L1}, \er{key1}, and \er{key2},
\begin{equation}
P_{\kp_n\ot  \sum_i c_i\cdot   \xi_{AA'}^{i1}\,\ot \xi_{BB'}^{i1}}(e_A^i|\lm)=P_{1_{H'''_A}\ot U_B (\ps_{AB}''')}(e_A^i|\lm),
\label{c2}
\end{equation}
and hence
\begin{equation}
P_{\psi}(Z=z_i|\lm)=P_{1_{H'''_A}\ot U_B (\ps_{AB}''')}(e_A^i|\lm).\label{c3}
\end{equation}
\item From \qm, notably \er{BR}, and \er{key3}, for any $i'\neq i$ we have
\begin{equation}
P_{1_{H'''_A}\ot U_B (\ps_{AB}''')}(e_A^{i'}\ot \xi_{BB'}^{ij_i})=0.\label{c4}
\end{equation}
\item From \textbf{CQ} and \er{c4}, for any $i'\neq i$,
\begin{equation}
P_{1_{H'''_A}\ot U_B (\ps_{AB}''')}(e_A^{i'}, \xi_{BB'}^{ij_i}|\lm)=0.\label{c5}
\end{equation}
\item From  \textbf{PI}, 
\begin{eqnarray}
P(e_A^{i'}|\lm)&=&\sum_{i,j_{i}}P(e_A^{i'},\xi_{BB'}^{ij_i} |\lm);\label{c6}\\
P(\xi_{BB'}^{ij_i} |\lm)&=&\sum_{i'} P(e_A^{i'},\xi_{BB'}^{ij_i} |\lm).\label{c7}
\end{eqnarray}
\item From \er{c5}, \er{c6}, and \er{c7},
\begin{equation}
P_{1_{H'''_A}\ot U_B (\ps_{AB}''')}(e_A^{i}|\lm)=\sum_{j_{i}}P_{1_{H'''_A}\ot U_B (\ps_{AB}''')}(\xi_{BB'}^{ij_i} |\lm).
\label{c8}
\end{equation}
\end{itemize}
Finally, from \er{c3}, \er{c8},   \er{q5a} - \er{q5b}, and \er{cmq} we obtain
\begin{equation}
P_{\psi}(Z=z_i|\lm) \stackrel{\varep}{\approx}\sum_{j_{i}}^{m_i} q^2=m_i\cdot q^2=(c'_i)^2.
\end{equation}
Since $c_i>0$ we have $c_i^2=|c_i|^2$; using \er{old4.36} and letting $\varep\raw 0$ then proves  step 3:
\begin{equation}
P_{\psi}(Z=z_i|\lm) =|c_i|^2 =P_{\psi}(Z=z_i).
\end{equation}

Finally, we remove our standing assumption that the spectrum of $Z$ be nondegenerate.
In the degenerate case one has 
\begin{equation}
P_{\psi}(Z=z_i)=\sum_{j_i}P_{\psi}(e_{j_i}),
\end{equation}
 where the sum is over any orthonormal basis $(e_{j_i})_{j_i}$ of the eigenspace of $z_i$. 
Since each state $e_{j_i}$ gives the same numerical outcome $Z=z_i$, probability theory gives for all $\lm$,
\begin{equation}
P(Z=z_i|\lm)=\sum_{j_i}P(e_{j_i}|\lm). 
\end{equation}
The nondegenerate case of the theorem (which distinguishes the states $e_{j_i}$) yields
 \begin{equation}
P_{\psi}(e_{j_i}|\lm)=P_{\psi}(e_{j_i}),
\end{equation}
from which  \er{CR1} follows once again: $$
P_{\psi}(Z=z_i|\lm)=\sum_{j_i}P_{\psi}(e_{j_i}|\lm)=\sum_{j_i}P_{\psi}(e_{j_i})=P_{\psi}(Z=z_i).$$
\appendix
\section{Embezzlement}
We only treat the amazing technique of embezzlement for maximally entangled states (cf.\ \cite{vDH} for the general case). We will deal with three \Hs s, namely
$H=\C^l$, $H'=\C^m$, and $H''=\C^n$ (where $n=m^N$ for some large $N$, see below), each with some fixed orthonormal basis 
 $(e_i)_{i=1}^l$,  $(e'_j)_{j=1}^m$, and $(e''_k)_{k=1}^n$, respectively. 
 Given a further number $m_i\leq m$, we now  list the  $nm$ basis vectors $e''_k\ot e'_j$ of $H''\ot H'$ in  two different orders:
\begin{enumerate}
\item $e''_1\ot e'_1,  \ldots, e''_n\ot e'_1, e''_1\ot e'_2,  \ldots, e''_n\ot e'_2, \ldots,  e''_1\ot e'_m, \ldots, e''_n\ot e'_m$;
\item $e''_1\ot e'_1,  \ldots, e''_1\ot e'_{m_i}, e''_2\ot e'_1,  \ldots, e''_2\ot e'_{m_i}, \ldots,  e''_n\ot e'_1, \ldots, e''_n\ot e'_{m_i}, \ldots$, 

\noindent where the remaining vectors (i.e., those of the form $e''_k\ot e'_j$ for $1\leq k\leq n$ and $j>m_i$) are listed in some arbitrary order. 
\end{enumerate}
Define  $U^{(m_i)}:H''\ot H'\raw H''\ot H'$ as the unitary operator that maps  the first list on the second. We will need the explicit expression
\begin{equation}
U^{(m_i)}(e''_k\ot e'_1)=e''_{s^i_k}\ot e'_{j^i_k}, \label{defUi}
\end{equation}
where for given $k=1, \ldots, n$  the numbers $s^i_k=1, \ldots, n_i$ (where $n_i$ is the smallest integer such that $n_im_i\geq n$) and $j^i_k=1, \ldots, n_i$ are
uniquely determined by the decomposition 
\beq k=(s_k^i-1)m_i+j^i_k.\eeq

We will actually work with two copies of $H''\ot H'$, called $H_A''\ot H_A'$ and $H_B''\ot H_B'$, with ensuing copies of $U_A^{(m_i)}$ and $U_B^{(m_i)}$
of $U^{(m_i)}$, and hence,  leaving the  isomorphism  $$H_A''\ot H_A'\ot H_B''\ot H_B'\cong H_A''\ot H_B''\ot H_A'\ot H_B'$$  implicit, we obtain a unitary operator
\beq
U_A^{(m_i)}\ot U_B^{(m_i)}:  H_A''\ot H_B''\ot H_A'\ot H_B'\raw H_A''\ot H_B''\ot H_A'\ot H_B'.\label{uAuB}
\eeq
The point of all this is that the unit vector $\kp_n\in H_A''\otimes H_A''$ defined by
\begin{equation}
\kp_n=\frac{1}{\sqrt{C(n)}}\sum_{k=1}^ne''_k\ot e''_k, \label{mun}
\end{equation}
where $C(n)=\sum_{k=1}^n 1/k$, acts as a catalyst in producing the maximally entangled state
\begin{equation}
\phv=\frac{1}{\sqrt{m_i}}\sum_{j=1}^{m_i} e_j'\ot e_j', \label{defmestate}
\end{equation}
in $H_A'\ot H_B'$ from the uncorrelated state $e_1'\ot e_1'\in H_A'\ot H_B'$, in that for any $m_i\leq m$,
\begin{equation}
U^{(m_i)}_A\ot U^{(m_i)}_B(\kp_n\ot e_1'\ot e_1')\stackrel{\varep/2}{\approx}\kp_n\ot \phv, \label{fe}
\end{equation}
where $\varep=1/N$ if $n=m^{2N}$. This follows straightforwardly from \er{uAuB} - \er{defmestate}.
\section*{Acknowledgement}
The author is indebted to Eric Cator, Roger Colbeck,  Dennis Dieks,  Dennis Hendrikx, and especially Gijs Leegwater for discussions, correspondence, and friendly corrections.

\end{document}